\documentclass[useAMS,usenatbib]{mn2e}

\usepackage{graphicx}
\usepackage{color}
\newcommand{\pd}[2]{\frac{\partial #1}{\partial #2}}
\newcommand{\av}[1]{\bar{#1}}
\renewcommand{\vec}[1]{\bmath{#1}}

\title[3D RMHD Simulations of Jets]{High Resolution 3D Relativistic MHD Simulations of Jets}
\author[Mignone et al.]{A. Mignone$^{1}$\thanks{E-mail:
mignone@ph.unito.it (AM)}, P. Rossi$^{2}$, G. Bodo$^{2}$, A. Ferrari$^{1}$
and S. Massaglia$^{1}$\\
$^{1}$Dipartimento di Fisica Generale dell Universit\'a di Torino, Italy\\
$^{2}$INAF Osservatorio Astronomico di Torino, Italy}

\begin{document}


\pagerange{\pageref{firstpage}--\pageref{lastpage}} \pubyear{2009}

\maketitle

\label{firstpage}

\begin{abstract}
  Relativistic magnetized jets are key elements in Active Galactic
  Nuclei and in other astrophysical environments. 
  Their structure and evolution involves a complex nonlinear physics 
  that can be approached by numerical studies only.
  Still, owing to a number of challenging computational aspects, only a few
  numerical investigations have been undertaken so far.
  In this paper, we present high-resolution three dimensional numerical 
  simulations of relativistic magnetized jets carrying an initially
  toroidal magnetic field. The presence of a substantial toroidal component 
  of the field is nowadays commonly invoked and held responsible for the 
  process of jet acceleration and collimation.
  We find that the typical nose cone structures, commonly observed
  in axisymmetric two-dimensional simulations, are not produced in
  the 3D case. Rather, the toroidal field gives rise to strong current driven kink
  instabilities leading to jet wiggling. However, it appears to be
  able to maintain an highly relativistic spine along its full length.
  By comparing low and high resolution simulations, we emphasize the 
  impact of resolution on the jet dynamical properties.
\end{abstract}


\section{Introduction}
%
%
%
%
%
%
%

Understanding the dynamics of relativistic jets is an essential step
for the comprehension and interpretation of many phenomenological
properties of Active Galactic Nuclei. Magnetic fields appear to be an
essential ingredient in their structure: observations of highly
polarized non-thermal synchrotron radiation evince the presence of
partially ordered magnetic fields, e.g., \cite{Gab04, Lain08, Gab08}.
In particular, it is likely that a toroidal component prevail on large 
scales as the poloidal component decays as $B_{p}\propto 1/R_j^2$ while
the toroidal $B_{\phi}\propto 1/R_j$, where $1/R_j$ is the jet radius that
expands from the inner galactic core to the radio jet lobes.
From the theoretical point of view toroidal magnetic fields are the 
necessary element of the widely accepted magnetically driven mechanisms 
for jet acceleration and collimation by fields with footpoints attached 
to a spinning black hole or accretion disc \citep{Lov76, Bland76, BP84, 
Narayan07}. 

On the other hand cylindrical magnetohydrodynamic (MHD) configurations 
are subject to many unstable modes, as reflection modes, Kelvin-Helmholtz 
modes, current carrying modes, etc. 
In particular configurations in which toroidal magnetic fields dominate 
are known to be violently unstable to the $m=1$ kink instability according
to the Kruskal-Shafranov criterion \citep{bate78}
$$ 
\left |{B_{\phi} \over B_p} \right | > {{2 \pi R_j} \over {z}}
$$
where $z$ is the jet length. 
Instead astrophysical jets appear to be quite stable and this becomes 
an unsolved issue, although several proposals have been made, related 
to the formation of cocoons by the bow shock at the head of the jet 
\citep{Begel98}, \citep{Mass96} or the jet expansion \citep{Spr08, MB09},
or wave coupling with MHD modes producing particle acceleration 
\citep{Benf80}.

On the other hand, one may wonder whether the main jet constituent is
electromagnetic energy in the form of Poynting-dominated beams, rather
than mass.  However, \cite{Sik05} argued that, even though jets could
be Poynting-dominated at the origin, observational data suggest that
they become kinetically dominated beyond about 1,000 gravitational
radii from the central acceleration region, powered by a super-massive
black-hole of $\sim 10^{8-10}$ M$_\odot$. 

The mechanism by which the toroidal field responsible for the jet 
acceleration can be dissipated may be related to the above mentioned 
instabilties. In this respect
\cite{Gian06} 
\citep[see also][]{Eichler93, Begel98} analyzed the role of the kink 
instability in magnetically driven jets and concluded that the Poynting 
flux can be rapidly dissipated and the jet becomes kinetically-dominated, 
again at about 1,000 gravitational radii. 

Recently, \citet{MB09} showed that relativistic 
jets produced in proximity of rapidly spinning black-holes with 
dipolar magnetic fields expand rapidly and can survive up to large 
radii without showing appreciable disruption due to 
the dominant $m=1$ kink mode.
Their simulations were carried out using large-scale 
global general relativistic simulations in three-dimensional 
spherical coordinates adopting a modest resolution to capture 
both the disk and the jet dynamics.
Three-dimensional simulations of the jet formation and evolution
in the general relativistic MHD domain has been performed as well by \citet{nish05}
in the case of a Schwarzschild black hole accretion disk system.

It is therefore crucial to study in a more systematic fashion the 
fate of relativistic magnetized jets in this transition region, 
examine the model parameter space to find whether the above requirements 
and predictions can be met. 
Relativistic jets carrying magnetic fields can be described by the 
equations of relativistic MHD (RMHD) and only in the last decade 
considerable efforts have been spent towards obtaining more 
robust and accurate numerical schemes. 
This opened the door to noticeable advancements in the field of 
numerical simulations. A first study of the properties
of two-dimensional relativistic magnetized jets has been performed by \citet{kom99}
and subsequently they have been thoroughly analyzed by
\citet{leis05, MMB05, KMHC08}. \citet{hardee07} have, instead, studied the
instability of a 3D RMHD cylindrical flow. 

In this paper we present
the first high-resolution (to our knowledge) numerical 
simulations of the propagation in three dimensions of a (special) 
relativistic magnetized jet injected in a uniform un-magnetized external 
medium. 
Computations are carried out at both high ($640\times 1600\times 640$ cell,
with $20$ pts/beam radius) and low ($320\times 800\times 320$, with $10$
pts/beam) resolution.  A comparison between the two reveals the
importance of high resolution, especially for this class of astrophysical
phenomena. We consider jets initially carrying a purely toroidal magnetic
field: as previously said, the presence of this component is
intimately connected to the jet formation mechanism and it is likely
to affect the jet dynamics through the onset of current driven (CD) 
instabilities. In this respect, it is of great interest to analyse their growth and
evolution. 
Previous non-relativistic three dimensional simulations of \cite{NM04} 
revealed that strongly magnetized supersonic jets develop CD induced 
distortions and that the kink ($m=1$) mode grows faster than the other modes.
More recently, \citet{Spr08} have studied the generation and
propagation of a nonrelativistic jet with a self-consistent toroidal
field component, showing and analysing the development of these
instabilities. In the present work we investigate the relativistic case,
starting with the limiting case of a purely toroidal field which,
at least in the Newtonian case, is known to be the most unstable 
configuration. Our study extends previous 3D simulations of the 
propagation of unmagnetized relativistic jets  \citep{Rossi08}
and, in the following, we will often refer to these results for
comparison.  
The effects of different field configurations and physical parameters 
will be examined in a forthcoming paper.

In the next section we describe our numerical setup, 
in \S\ref{sec:results} we outline the simulation results and
in \S\ref{sec:conclusions} we summarize our conclusions.

\section{Numerical Setup}
\label{sec:numsetup}
%
%
%
%

Numerical simulations are carried out by solving the equations 
of special relativistic MHD (RMHD) in conservation form \citep{Anile90}:
\begin{equation}\begin{array}{lcl}
  \label{eq:clmass}
  \partial_\mu\left(\rho u^\mu\right) & = &  0 \,, \\ \noalign{\medskip}
  \label{eq:clmomen}
  \partial_\mu\Big(w u^\mu u^\nu - b^\mu b^\nu + p\eta^{\mu\nu}\Big) & = & 0\,,\\ \noalign{\medskip}
  \label{eq:clind}
  \partial_\mu\left(u^\mu b^\nu - u^\nu b^\mu\right) & = & 0\,,
\end{array}
\end{equation}
where $\rho$ is the rest mass density, $u^\mu\equiv \gamma(1,\vec{v})$ 
and $b^\mu = (b^0, \vec{B}/\gamma + b^0\vec{v})$ are the four-velocity
and covariant magnetic field written in terms of the three velocity 
$\vec{v}$ and laboratory magnetic field $\vec{B}$ and
$b^0=\gamma\vec{v}\cdot\vec{B}$.
A flat metric $\eta^{\mu\nu}={\rm diag}(-1,1,1,1)$ is considered.
In the previous equations $w = w_{\rm g} + b^2$ and
$p = p_g + b^2/2$ express the total enthalpy and total pressure
in terms of their thermal ($w_{\rm g}$ and $p_{\rm g}$) and 
magnetic contributions ($b^2 = b^\mu b_\mu$), respectively.
An additional equation, describing the advection of a passive scalar 
(or tracer) $f(x,y,z,t)$, is included to discriminate
between jet material (where $f=1$) and the environment (where $f=0$). 
We assume a single-specie relativistic perfect fluid (the 
Synge gas) described by the approximated equation of state proposed 
by \cite{MPB05} and \cite{MMcK07}.

The computational domain is initially filled with an un-magnetized
medium at rest, with uniform density $\rho_a$ and gas pressure $p_a$.
The jet flows parallel to the $y$ direction through a cylindrical 
nozzle  specified by $r = \sqrt{x^2 + z^2} < 1$ 
(in units of the jet radius) at the lower boundary.  
At the nozzle, the beam has constant density $\rho_j$ and longitudinal 
velocity component ($v_y$) given in terms of the Lorentz factor $\gamma_j$, whereas
the magnetic field carries a purely toroidal (i.e. azimuthal) component: 
\begin{equation}\label{eq:btor}
  B_\phi(r) = \left\{\begin{array}{ll}
  B_m r/a & \quad\textrm{for}\quad r < a \,,\\ \noalign{\medskip}
  B_m a/r & \quad\textrm{for}\quad r > a \,,\\ \noalign{\medskip}
   0      & \quad\textrm{for}\quad r > 1 \,,
\end{array}\right.
\end{equation}
where $a = 1/2$ is the magnetization radius.
The pressure profile follows from the numerical integration of 
the radial momentum balance equation across the beam:
\begin{equation}\label{eq:p_balance}
\pd{}{r}\left(p_g + \frac{b^2}{2}\right) = \frac{wu_\phi^2 - b_\phi^2}{r}\,.
\end{equation}
where a small differential rotation is assumed by setting
$u_\phi = \alpha\gamma B_\phi/B_m$ with $\alpha = 0.2$. 
The pressure profile $p(r)$ is continuous across the jet 
boundary, i.e., $p_a = p_j(1)$.
The integration constant and the field strength $B_m$ are specified 
through the sonic Mach number $M_s = v_j/c_s(a)$ 
and Alfv\'enic Mach number $M_A = v_j/v_A(a)$ ($c_s$ and $v_A$ are the sound 
and Alfv\'en velocities) at the magnetization radius. 
Thus the beam values are entirely specified in terms
of the density contrast $\eta = \rho_j/\rho_a$, Lorentz factor $\gamma_j$,
sonic and Alfv\'enic Mach numbers.
Motivated by the results obtained in \cite{Rossi08}, we choose 
$\eta = 10^{-4}$, $\gamma_j = 10$, $M_s = 3$ and $M_A = 1.67$,
so that the ratio of the average gas and magnetic pressures is close to one
(equipartition).
Velocity perturbations are super-imposed as in \cite{Rossi08}.
Reflective boundary conditions hold at $y=0$ outside the inlet region 
and zero-gradient is assumed through the remaining boundaries.

Simulations are carried out with the PLUTO code for 
astrophysical gasdynamics \citep{PLUTO}. 
The chosen configuration employs a second order Godunov type 
scheme with the recently developed HLLD Riemann solver 
\citep{MUB09} and the mixed hyperbolic/parabolic divergence cleaning by
\cite{Dedner02} to control the solenoidal constraint.
Our computational box is defined by 
$x\in [-L/2,L/2]$, $y\in [0,L_y]$ and $z\in [-L/2,L/2]$,
($L = 56$, $L_y = 80$) with a uniform resolution 
in $|x|,|z| < 10$ and a geometrically stretched grid
otherwise.
Both low ($10$ points per beam radius) and high 
($20$ zones) resolution simulations are performed,
yielding an overall resolution of
$320\times 800 \times 320$ and $640\times 1600\times 640$
computational zones, respectively.

\section{Results}
\label{sec:results}
%
%
%
%

\begin{figure*}
\includegraphics[width=0.45\textwidth]{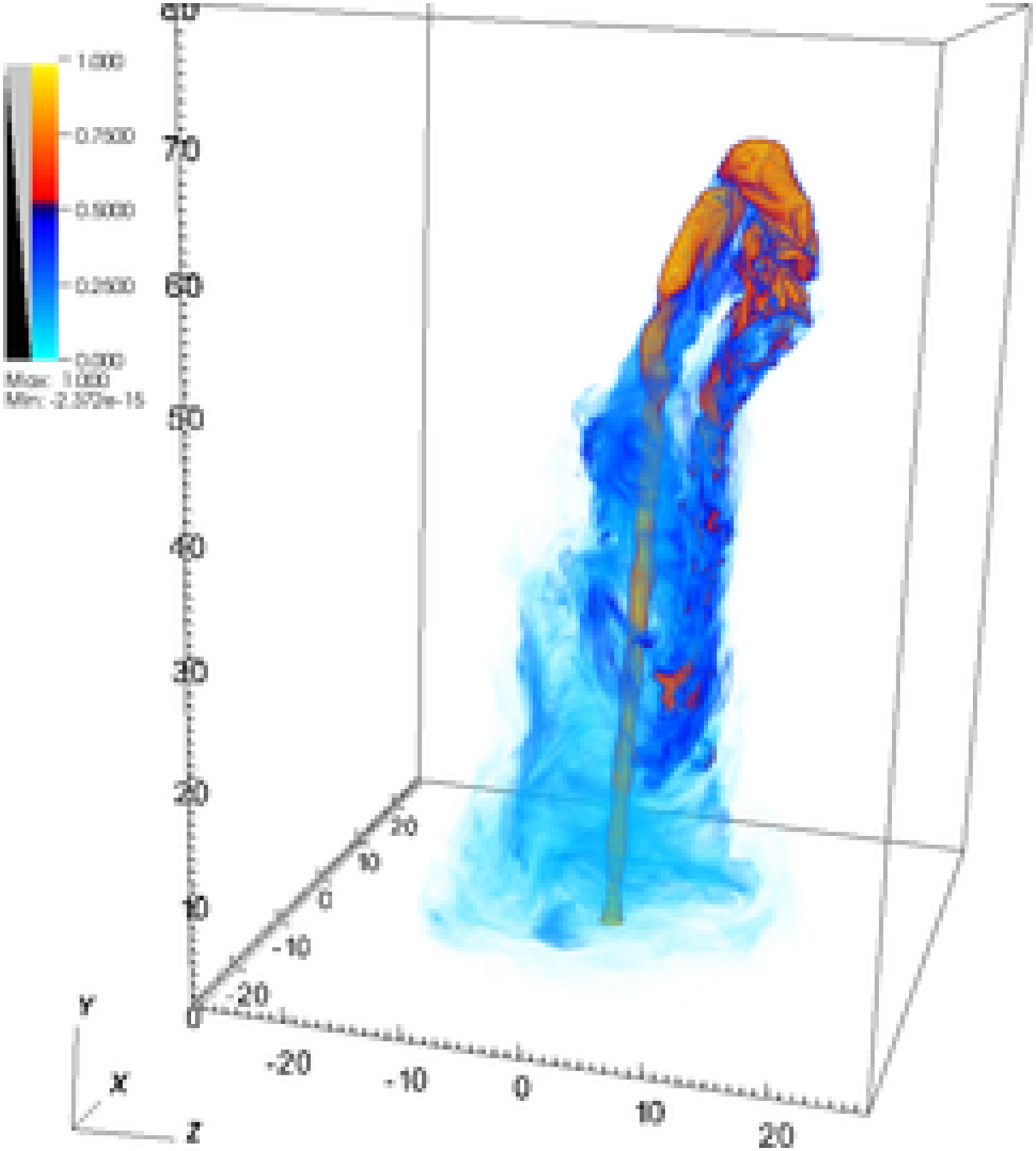}%
\includegraphics[width=0.45\textwidth]{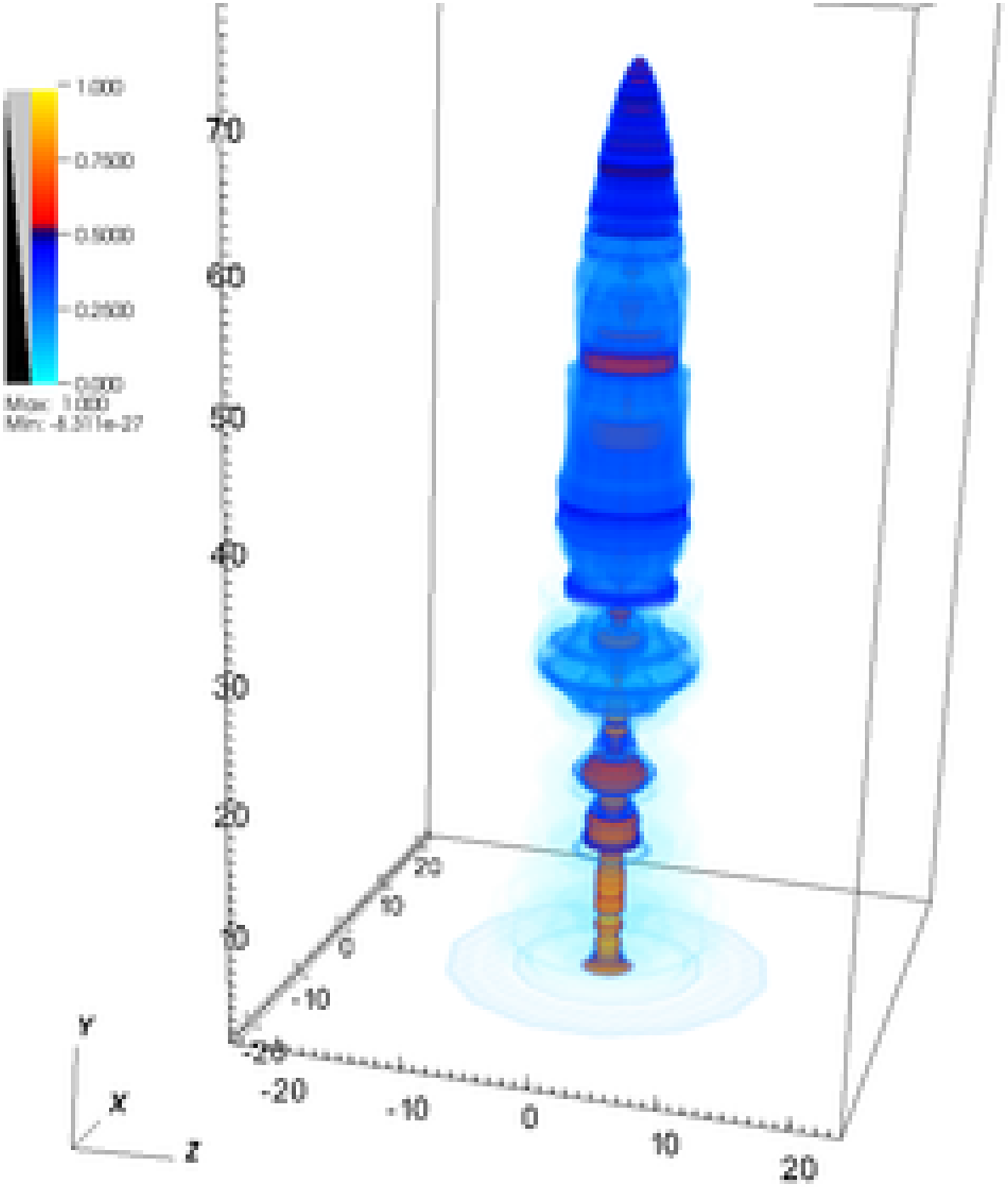}
\caption{Volume renderings of the passive scalar distributions for 
         the high resolution 3D run (left panel) and 2D axisymmetric case 
         (right) at $t=710$.}
\label{fig:2d3d}
\end{figure*}%
\begin{figure}
\includegraphics[width=0.45\textwidth]{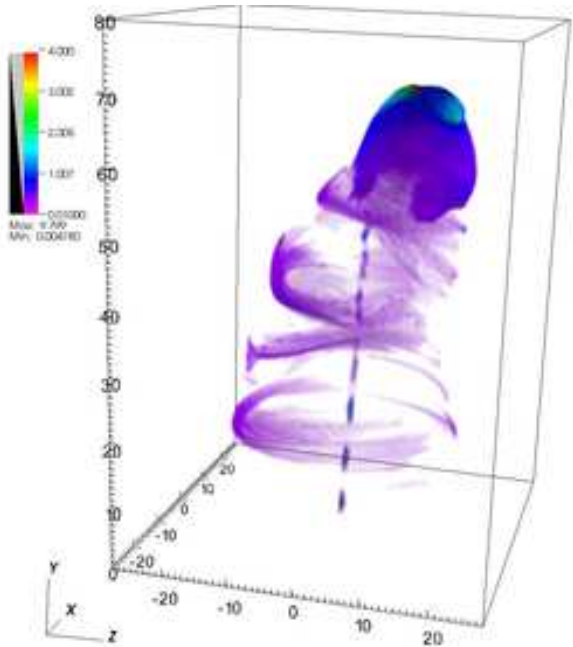}
\caption{Volume renderings of the thermal pressure distributions at 
         $t=710$ for the high resolution 3D run.}
\label{fig:3d_press}
\end{figure}

In the left panel of Fig. \ref{fig:2d3d} we show the volume renderings of the 
tracer distribution when the jet has reached the length of $\sim 70$ 
beam radii. 
For the sake of comparison we also show, in the right panel, the rendering 
of an equivalent 2-D axisymmetric simulation with the same parameters.
The picture (in the right panel) clearly indicates the presence of a
nose cone structure typical of two-dimensional high Poynting flux jets
\citep{CNB86, Bla89, kom99, leis05}, originating from  
the amplification at the terminal shock  of the toroidal field component which 
then confines the jet material and prevents it to freely flow in the cocoon.
In three dimensions, however, this structure becomes unstable and does not form leading 
to a very different asymmetric morphology. In following the jet propagation, we 
observe that its trajectory becomes progressively more curved, moving away from the
longitudinal $y$ axis. This effect becomes more pronounced at the jet head
and can be attributed to the presence of current driven kink instabilities. 
Moreover, the backflowing material forms a very asymmetric cocoon as a 
result of the changes in direction of the jet head. 
In Fig. \ref{fig:3d_press}, representing the
pressure distribution, it is possible to observe two regions of
enhanced pressure in the bow shock. Comparing the tracer and
pressure distributions, we can identify these areas as the points where
the jet is more strongly deflected. The asymmetric backflow appears to
be quite pronounced, reaching relatively high velocities  up to $0.9 c$.

\begin{figure}\label{fig:com}
\includegraphics[width=0.45\textwidth]{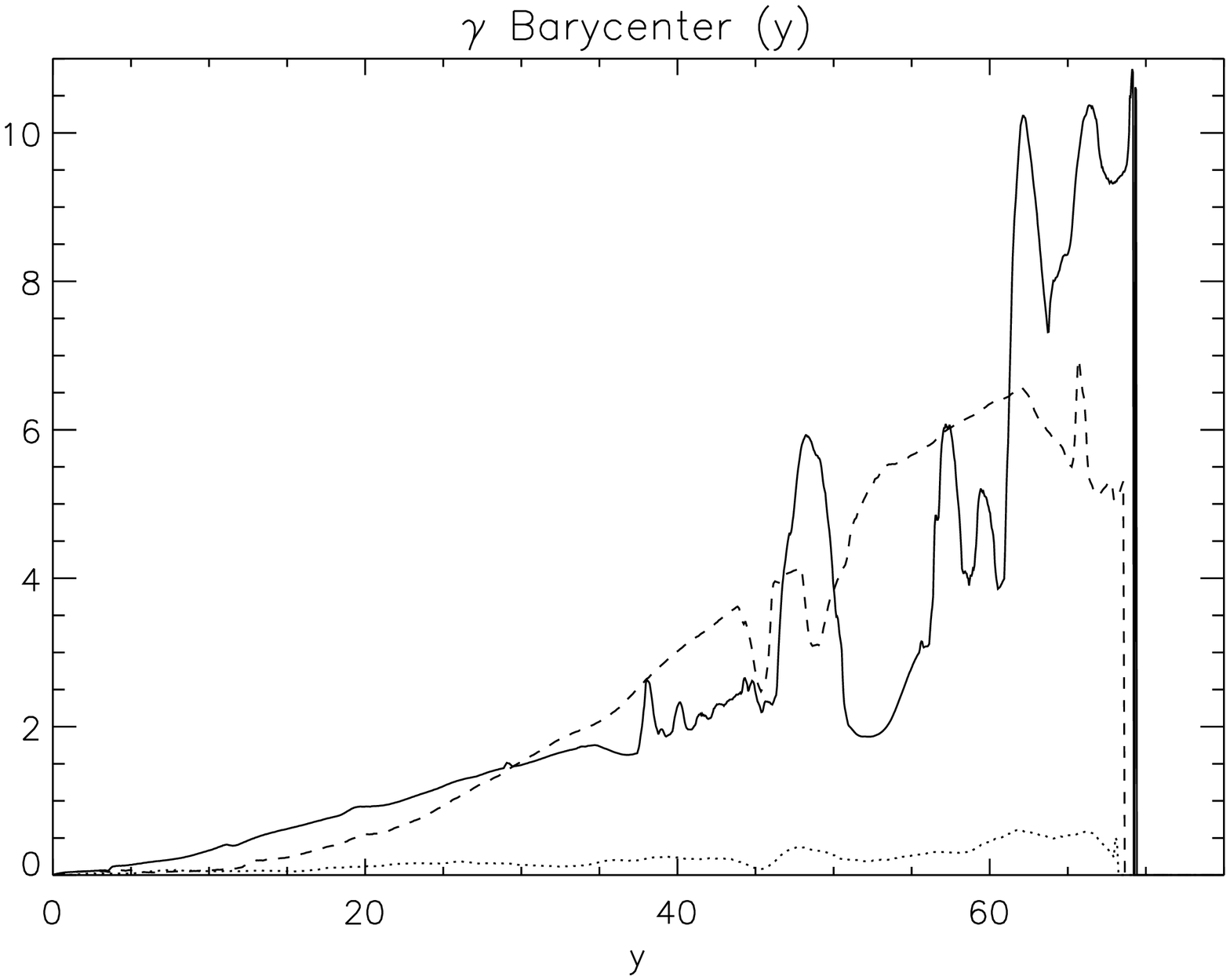}\\
\includegraphics[width=0.45\textwidth]{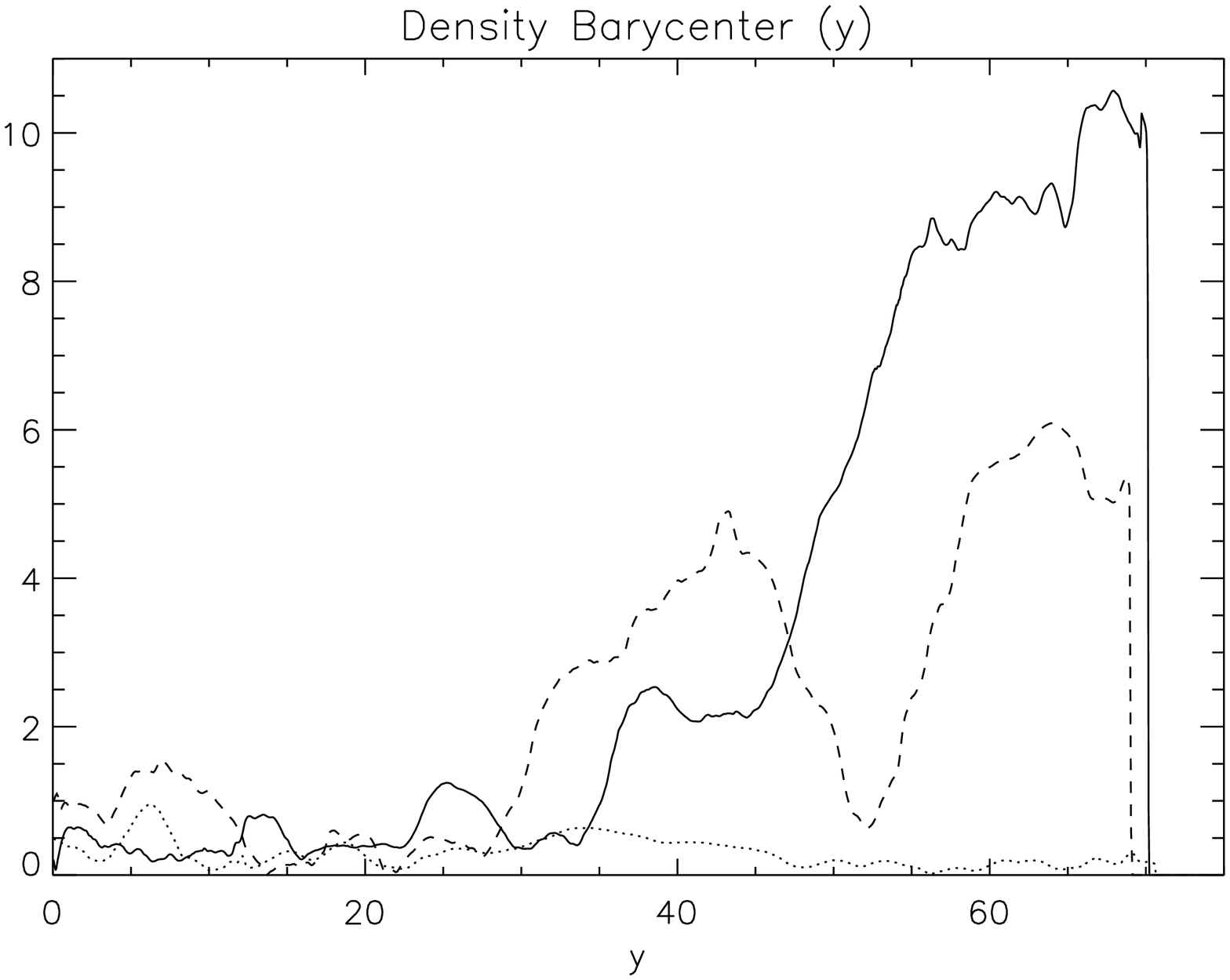}
\caption{Barycenter Displacements from the longitudinal axis
         $\av{r}$ of the Lorentz factor (top) and jet laboratory density 
         (bottom) for the high resolution (solid lines), low 
         resolution (dashed lines) and hydro (dotted lines) cases.
         The corresponding integration times are $t=710$, $t=840$
         and $t=590$, respectively.}
\end{figure}
The displacement of the beam from the longitudinal axis plays a fundamental
role in the jet morphology and it is evaluated, at each position along the
jet, by the quantity $\av{r}(y) = \sqrt{\av{x}^2 + \av{z}^2}$, where
\begin{equation} \label{eq:com}
 \av{x}(y) = \frac{\int x Q dx\,dz  }{\int Q dx\,dz}  \,,\quad
 \av{z}(y) = \frac{\int z Q dx\,dz  }{\int Q dx\,dz} \,.
\end{equation}
where $Q$ can be any flow quantity.
Choosing $Q=\gamma\chi$, where $\chi = 1$ for $\gamma > 1.5$ and $\chi=0$ otherwise,
gives a measure of the distance of the Lorentz factor 
centroid from the axis. Likewise, we can quantify the jet density
barycenter by choosing $Q = \rho\gamma f$. 
In Fig. \ref{fig:com} we plot both the Lorentz factor and density 
centroids as functions of $y$. Solid, dashed and dotted lines refer,
respectively, to the high resolution, low resolution and non magnetized case
with the same parameters (Lorentz factor, density ratio and Mach numbers).
Simulation times have been chosen so that the jets have reached approximately
the same distance, $\sim 70$ radii.
Both the low and high resolution cases show moderate displacements 
up to $\sim 40$ jet radii followed by a strong increase soon after. 
The effect is more pronounced in the high resolution case, eventually
reaching maximum values in excess of $10$. On the contrary, 
no significant deviations are seen in the purely hydro case and 
the beam propagates very close to the initial axis.

\begin{figure}
\includegraphics[width=0.45\textwidth]{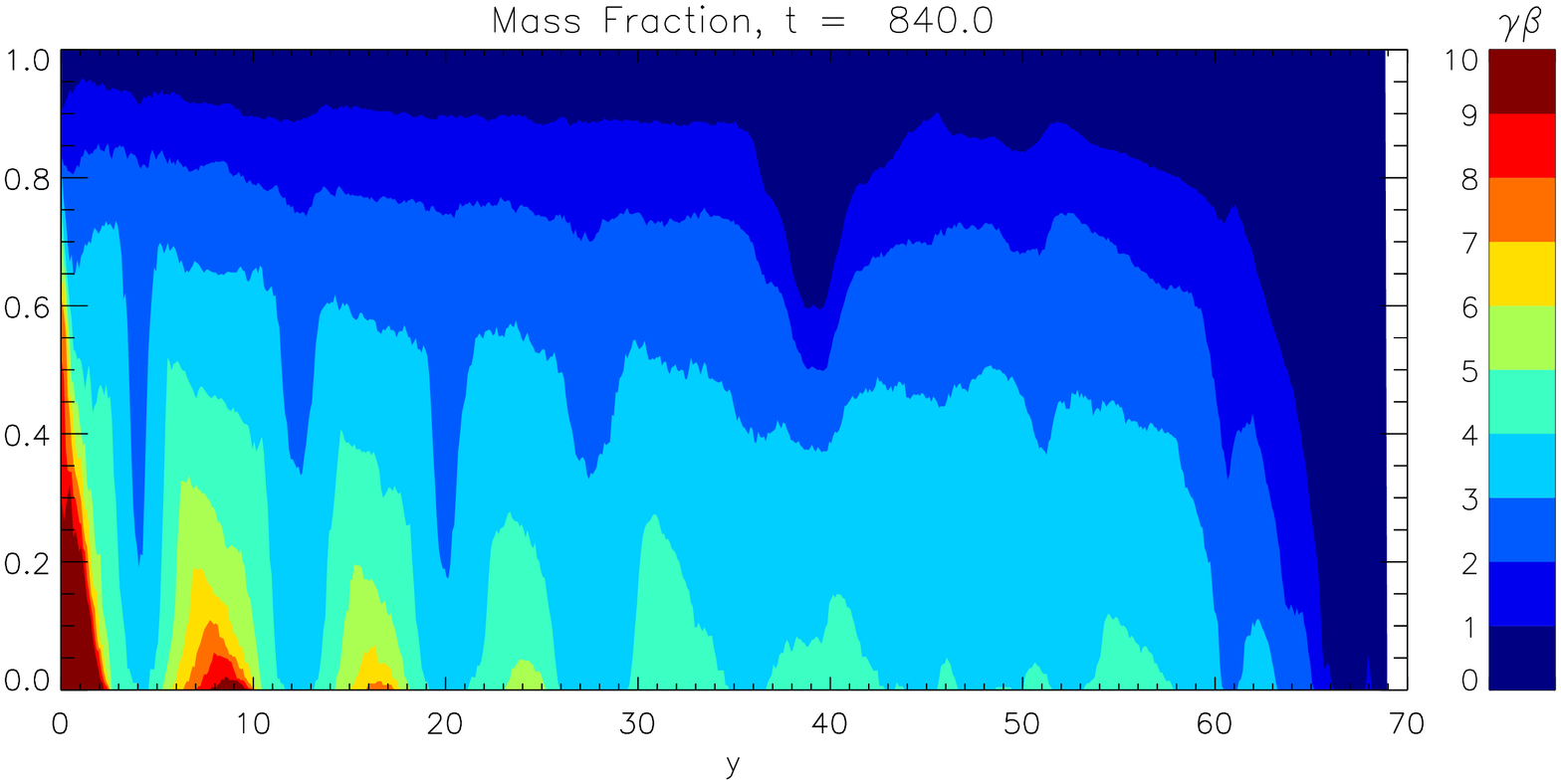}
\includegraphics[width=0.45\textwidth]{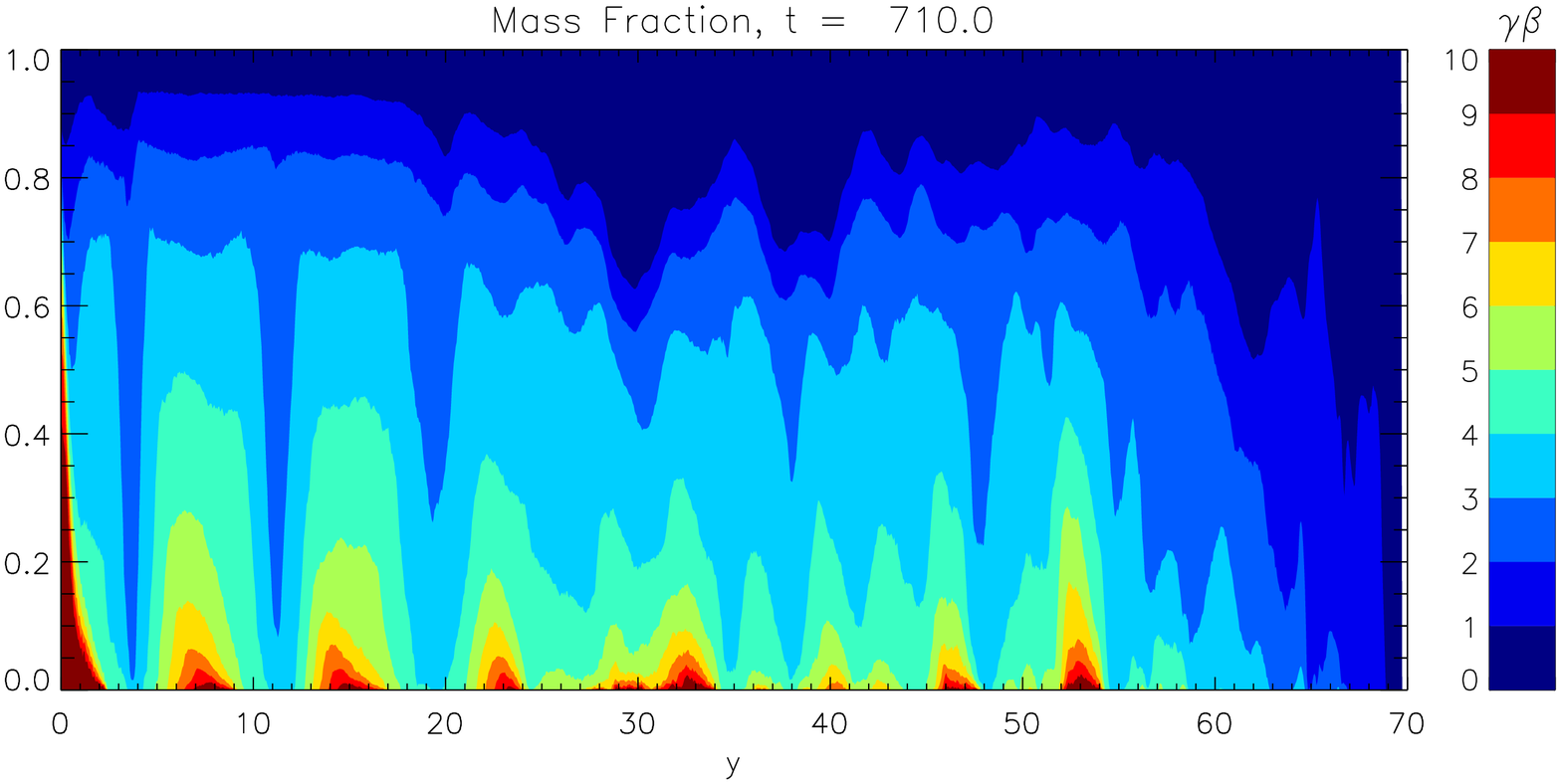}
\caption{Jet mass fraction for the low resolution (top panel)
         and high resolution (bottom panel) cases.}
\label{fig:entr}
\end{figure}
Fig. \ref{fig:entr} represents the distribution of the jet mass fraction 
moving at a certain value of $\gamma \beta$ at the end of the
simulation when the jet has traveled approximately $70$ beam radii. 
Both low (top panel) and high (bottom panel) resolution cases are shown. 
We use the four-velocity (instead of the Lorentz factor)
to avoid compression of the scale close to $\gamma = 1$, i.e. at low
velocities. The legend on the right gives the corresponding value
of $\gamma \beta$ for each color. As an illustrative example, 
one can see that at $y=15$ the mass fractions moving with $\gamma\beta > 6$,
$\gamma\beta\in[5,6]$,$\gamma\beta\in[4,5]$,$\gamma\beta\in[3,4]$ and
$\gamma\beta < 3$ are, respectively, $10\%$, $10\%$, $20\%$, $25\%$
and $35\%$.
Focusing on the high velocity part of the 
distribution, we can observe the presence of material moving at
$\gamma \sim 10$ all along the jet (red spots close to the y-axis). 
The behavior of the low resolution case, shown in the upper panel, is
quite different. Jet material moves at $\gamma \sim 10$ only up to $y
\sim 20$, with a decrease of the maximum Lorentz factor to a value of about
$5$. 

\begin{figure*}
\includegraphics[width=0.8\textwidth]{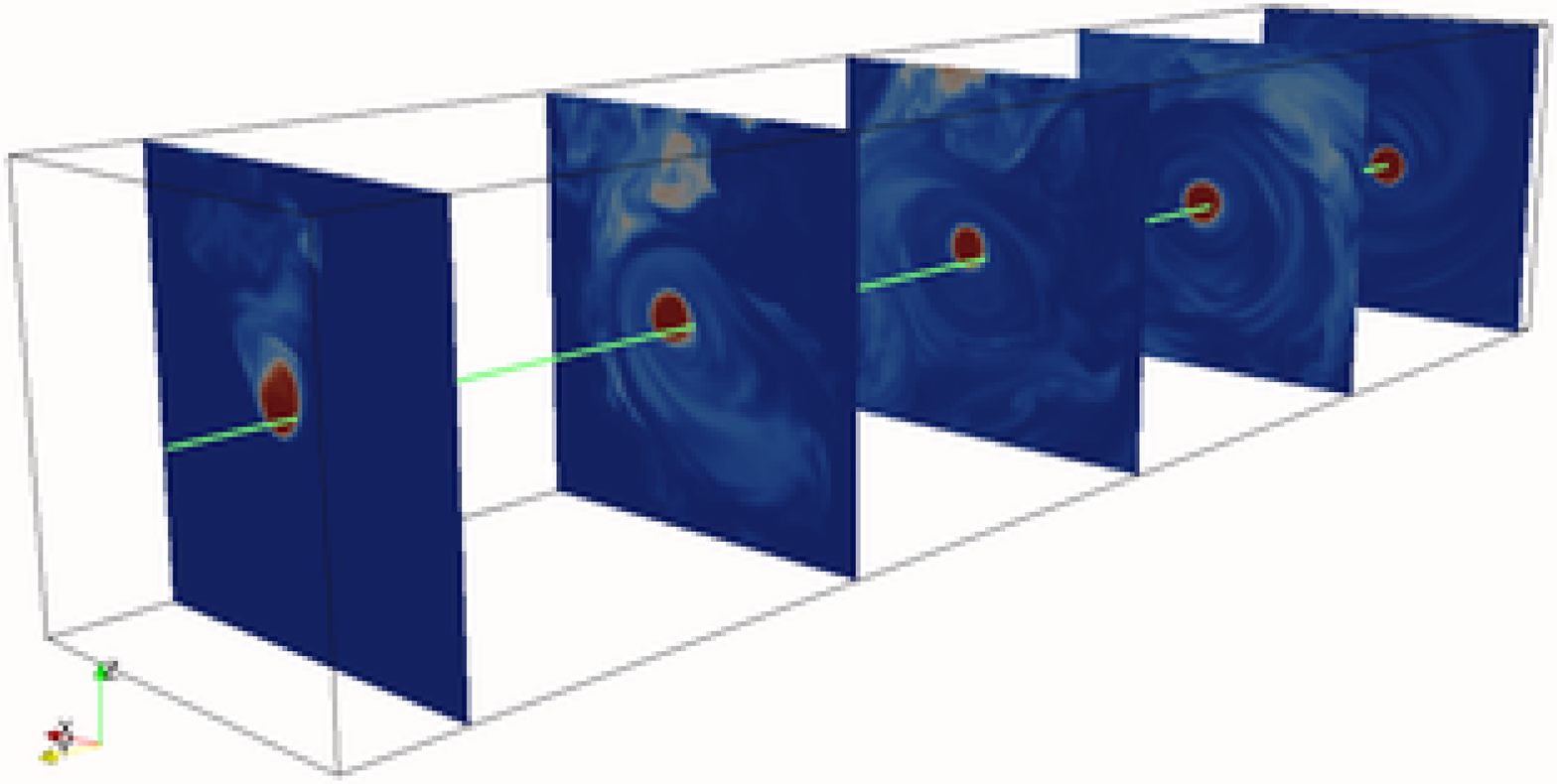}
\includegraphics[width=0.8\textwidth]{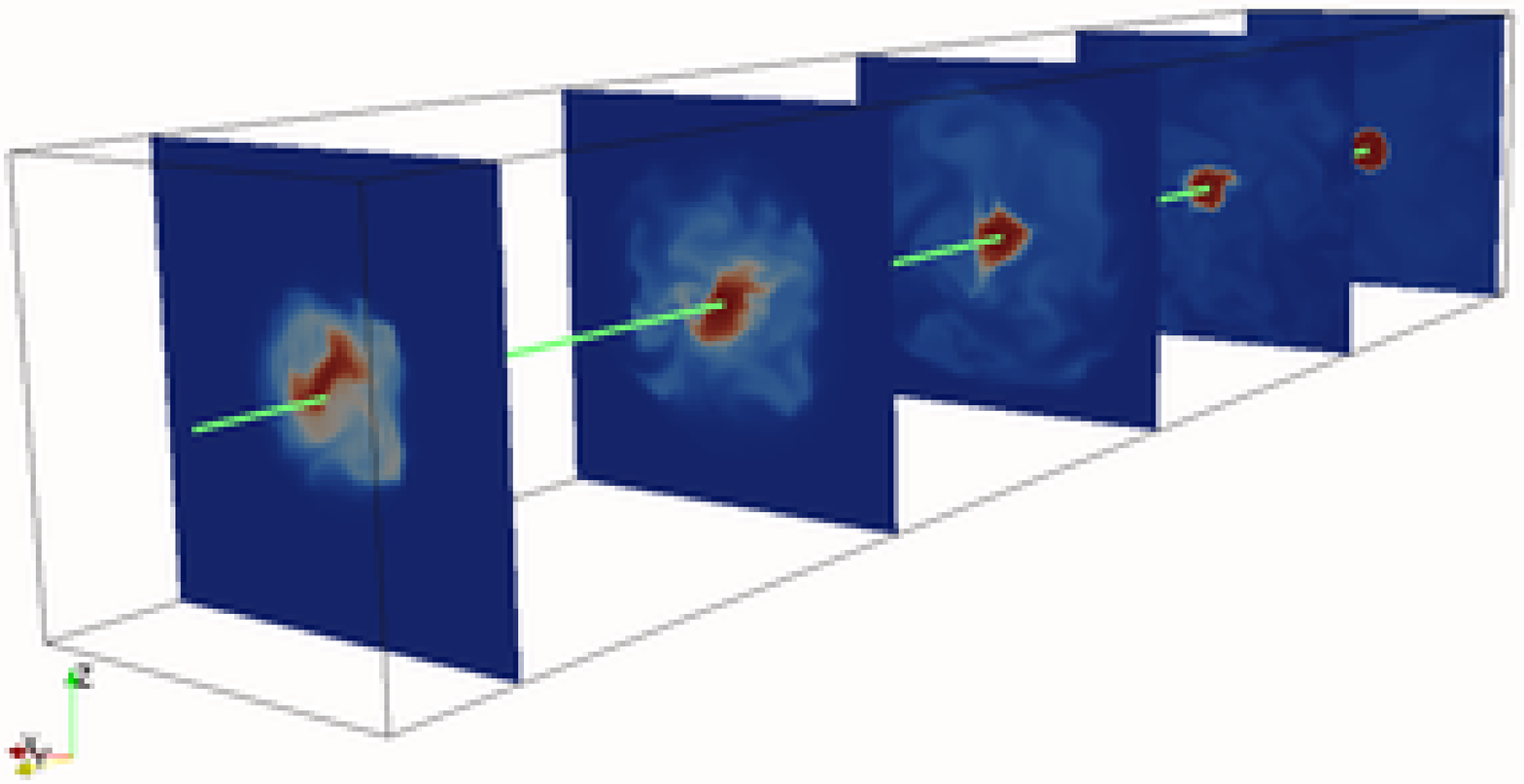}
\caption{Two dimensional slices at constant $y$ planes for the  
         high resolution magnetized case (top panel) and purely 
         hydrodynamical case (bottom, low resolution). The solid line marks the 
         longitudinal axis ($x,z=0$) while the line of sight is such 
         that the jet propagates from the furthest boundary to 
         the closer one.}
\label{fig:slices}
\end{figure*}
The toroidal field seems to protect the jet core from any 
interaction with the surrounding, thus preventing 
momentum transfer to the external medium. 
This behavior is better understood by inspecting the top
panel in Fig. \ref{fig:slices} where we show the jet cross-section 
(in the high resolution case) at different positions along the longitudinal axis. 
For comparison, a similar plot for the unmagnetized case, shown 
in the bottom panel, reveals that in this case the jet surface is 
deformed by non-axisymmetric modes of high order $(m > 2)$ promoting 
mixing between jet material and external medium.
On the contrary, the high resolution magnetized case shows the presence 
of low order modes $(m \leq 2)$ only, producing oval deformations that cannot
induce any mixing. The tension of the toroidal field, in fact, acts
as a strong stabilizing factor for high order modes.
Our results are thus in agreement with those of \cite{RHCJ99} where the presence
of a toroidal field was shown to noticeably lessen the growth and amplitude
of small wavelength modes. 

From Fig. \ref{fig:entr} one notices also the periodic decrease of the maximum
Lorentz factor associated with jet pinching and the formation of
periodic shocks along the beam, also visible in the left panel of 
Fig. \ref{fig:3d_press}, where it is possible to follow the appearance 
of periodic compressions all along the jet.
The formation of these internal shocks has been observed in all jet 
simulations and is in general attributed to the interaction of the 
cocoon with the jet proper and to Kelvin-Helmholtz instabilities. 
Three-dimensional effects seemed to decrease the importance 
of these internal shocks. Indeed, in a similar hydro case \citep{Rossi08} at
the same resolution, we observed their presence only in the first
half of the jet. Here, the decreased interaction of the jet core with 
the ambient favors their formation and, besides, the additional pinching
effect of the toroidal field increases the compression factor. 
At the same time, the presence of the toroidal field has also the effect of 
reducing the distance between the shocks when compared to the hydro case. 
We have seen that, in the case that we are examining, the jet core is
protected by the toroidal field and maintains its initial Lorentz
factor all along its length. However, it is interesting to ask whether 
an effect of mixing and entrainment can actually take place 
on the more external jet layers. 
From Fig. \ref{fig:entr} we can observe that the mass fraction
of the material with $\gamma \beta < 2$ varies between $20\%$ in the 
region closer to the jet origin, to about $30\%$ at outer positions, 
while in the hydro case this fraction reached values of more than $50\%$ \citep{Rossi08}, 
showing a smaller entrainment efficiency in this case. 
Correspondingly, a larger fraction of the jet material moves
at higher values of $\gamma \beta$.

These results show that, in spite of the large deflection induced 
by the presence of strong toroidal magnetic field, non-axisymmetric 
$m=1$ screw modes do not lead to appreciable jet disruption.
These findings agree with those reported in \citet{MB09} where 
the $m=1$ modes are effective only in modeling the jet 
substructure without affecting the large-scale dynamics.
Although our simulation setup is noticeably different, this
qualitative agreement supports the general idea that 
some other intervening mechanism must act so as to 
suppress the instability growth, \citep{Spr08,NLT09}.

The simulation we are discussing has been performed at
high resolution employing $20$ grid points over the beam
radius. Numerical resolution effects are known to be important in jet
simulations and, for this reason, we have compared the high resolution
computation with a lower resolution one, with only $10$ points on the
jet radius. In the low resolution case, as it is shown in the upper panel
 of Fig. \ref{fig:entr}, numerical diffusion effects
tend to increase the deposition of momentum from the jet core to the
ambient and therefore the jet is not able to maintain its initial
Lorentz factor up to the head. In addition, we have also observed
 that periodic shocks along
the jet appear to be substantially weaker and they are not present
all along the jet as in the high resolution case.
Besides, as discussed above, the jet displacement
results larger at high resolution. 
This comparison shows that the differences due to grid resolution are
substantial and demonstrates the importance of performing high
resolution simulations for reproducing more accurately the 
physical behavior. 
A question that may occur is obviously whether the resolution
employed in this work is sufficient, since our results show that we 
have not yet reached convergence in the basic jet properties (a 
complete convergence for this problem cannot be reached). 
We have shown a tendency towards larger displacements of the jet 
off the axis as the resolution is increased, although this effect 
is, in any case, localized at the jet's head.
Still, we observe a reduction of the momentum transfer 
from the inner jet to the ambient as the numerical resolution increases.
We expect that these well defined trends in the physical characters 
of the solution should be maintained by a further increase in resolution, 
that may be feasible in future studies on this subject.
The effects described above may therefore change quantitatively, but the 
qualitative features that we outlined should remain similar to what 
we obtain. 

The results pertaining to the hydrodynamical cases discussed in \cite{Rossi08} 
indicated the density ratio $\eta$ as the main parameter governing the 
interaction between the beam and the external medium.
In order to investigate the role of $\eta$ for the present magnetized model, 
we performed a low resolution computation with $\eta = 100$. 
The results show that, up to the same length, the jet propagates almost undisturbed, 
without any sign of kink instability nor of mixing with the ambient medium.
We think that the most likely explanation for the difference between the two
cases at different density ratios (this parameter has no direct effect on the
kink instability)  should be related to the different advance velocity of
the jet head, that is much larger for $\eta = 100$ than for $\eta = 10,000$. In fact, 
the unstable perturbations are advected by the jet as they grow. In the case with $\eta = 10,000$
the jet head advances at a velocity that is much lower than the jet velocity and, therefore,
the unstable perturbations accumulate at the jet head and have enough time for their growth 
and can give origin to the jet deformation. In the lower ratio case, instead, the jet head advances
at a much higher velocity and the kink instability does not have enough time to grow. In this
case its effect should be visible only on a larger scale.

\section{Conclusions}\label{sec:conclusions}
%
%
%
%

We have presented the first results of a three-dimensional high-resolution numerical 
study of the propagation of relativistic magnetized jets injected into a uniform medium 
using the PLUTO code. 
At the highest resolution, the computations employed $20$ points on the 
jet radius for an overall resolution of $640 \times 1600 \times 640$ zones 
covering a Cartesian domain of $56 \times 80 \times 56$, in units of the beam radius.

From the variety of possible magnetic field configurations, we decided to start our 
study by considering jets initially carrying a purely toroidal magnetic field.
Indeed, the most accepted jet formation mechanisms predict that this field component 
has to be present and this strengthen the necessity of understanding its 
key role on the jet dynamics.
The presence of a toroidal component of the field is known to drive
current-driven kink instabilities which in our case are responsible for 
jet wiggling and beam deflection off the main longitudinal axis \citep{NLT09}.
%
%
The deviations tend to become more prominent towards the final parts of 
the jet, where we measured off-axis displacements in excess of $10$ radii over a 
length of $70$ radii.
The wandering of the jet head, induced by kink instability effects,
may create multiple sites where the jet impacts on the external medium
forming strong shocks. 
This behavior may originate the multiple  hot spots that are observed 
in several radiogalaxies \citep[see e.g.][]{Leahy97, Hardcastle07}. The asymmetry of the
backflow is another feature that replicates the observational
appearance of several objects. 

An additional effect of the toroidal field component seems that of 
shielding the inner jet core from any interaction with the 
surroundings and therefore from a loss of momentum. 
Furthermore, we have observed that the jet remains highly
relativistic all along its length. 
Quasi periodic shocks, formed by the combination of several 
effects like the interaction with the cocoon
and the toroidal field pinching, can be observed all along the jet extension. 
A spine {\it plus} sheath layer structure with shocks supports the idea 
that jets at sub-kpc scale can originate correlated variability at radio 
frequencies and in X and gamma rays (see for instance recent observations by the 
AGILE and FERMI missions and the TeV ground arrays \cite{donna09, veritas09}). 
The very high frequency emission would originate in the slower sheath layer by 
inverse Compton boosting of the optical and X photons produced in the relativistic spine. 
The X ray emission would be synchrotron emission not self-absorbed and would have
variability strictly connected to the gamma ray emission.
The radio emission would come from synchrotron emission in an outer region of 
the jet sheath \citep{tave08}.

Recently \citet{Spr08} have performed 3D simulations of a Newtonian
magnetized jet with a toroidal component and have shown the development of
kink instabilities that produce helical jet deformations. 
In their setup the jet is generated by maintaining a rotation
profile in the boundary zones and a spherical grid, with a non uniform spacing in the
radial direction, is used to describe the jet propagation.

The comparison between low and high resolutions cases 
shows significant differences in the formation of shocks along the beam.
In the high resolution simulations, shocks appear to be stronger, closer 
and in a larger number. In addition, the jet is able to keep a 
highly relativistic core for a longer distance.

The results presented here represent only a first step in
the analysis of the behavior of relativistic magnetized jets. We have
shown that three dimensional and high resolution effects are important
factors in this investigation. In a forthcoming paper 
we intend to explore the influence of different (e.g. poloidal 
or helical) field configurations.

\section*{Acknowledgments}

The numerical calculations were performed at CINECA in 
Bologna, Italy, thanks to INAF.

\label{lastpage}

\end{document}